\documentclass[aps, prd, reprint, superscriptaddress, floatfix, nofootinbib, amsmath, amssymb]{revtex4-2}

\usepackage[utf8]{inputenc}
\usepackage{graphicx}
\usepackage{physics}
\usepackage[dvipsnames]{xcolor}
\usepackage[colorlinks=true]{hyperref}
\hypersetup{urlcolor=BlueViolet, citecolor=Plum, linkcolor=PineGreen}
\usepackage{cleveref}

\begin{document}


\title{Optomechanical dark matter instrument for direct detection}

\author{Christopher G. Baker}
\email{c.baker3@uq.edu.au}
\affiliation{ARC Centre of Excellence for Engineered Quantum Systems, University of Queensland, St Lucia, Brisbane, 4072, QLD, Australia}

\author{Warwick P. Bowen}
\email{w.bowen@uq.edu.au}
\affiliation{ARC Centre of Excellence for Engineered Quantum Systems, University of Queensland, St Lucia, Brisbane, 4072, QLD, Australia}

\author{Peter Cox}
\email{peter.cox@unimelb.edu.au}
\affiliation{ARC Centre of Excellence for Dark Matter Particle Physics, \\School of Physics, The University of Melbourne, Victoria 3010, Australia}

\author{Matthew J. Dolan}
\email{dolan@unimelb.edu.au}
\affiliation{ARC Centre of Excellence for Dark Matter Particle Physics, \\School of Physics, The University of Melbourne, Victoria 3010, Australia}

\author{Maxim Goryachev}
\email{maxim.goryachev@uwa.edu.au}
\affiliation{ARC Centre of Excellence for Engineered Quantum Systems,  Department  of  Physics, University of Western Australia, Crawley 6009, Australia}

\author{Glen Harris}
\email{g.harris2@uq.edu.au}
\affiliation{ARC Centre of Excellence for Engineered Quantum Systems, University of Queensland, St Lucia, Brisbane, 4072, QLD, Australia}


\begin{abstract}
We propose the Optomechanical Dark-matter INstrument (ODIN), based on a new method for the direct detection of low-mass dark matter. We consider dark matter interacting with superfluid helium in an optomechanical cavity. Using an effective field theory, we calculate the rate at which dark matter scatters off phonons in a highly populated, driven acoustic mode of the cavity. This scattering process deposits a phonon into a second acoustic mode in its ground state. The deposited phonon ($\mu$eV range) is then converted to a photon (eV range) via an optomechanical interaction with a pump laser. This photon can be efficiently detected, providing a means to sensitively probe keV scale dark matter. We provide realistic estimates of the backgrounds and discuss the technical challenges associated with such an experiment. We calculate projected limits on dark matter--nucleon interactions for dark matter masses ranging from 0.5 to 300\,keV and estimate that a future device could probe cross-sections as low as $\mathcal{O} \left( 10^{-32}\right)\,\rm{cm}^2$.
\end{abstract}

\maketitle

\section{Introduction}

Cosmological and astrophysical observations provide strong evidence for the existence of dark matter~\cite{DM_Review_2018}. Understanding its fundamental nature remains one of the most important problems in particle physics, astrophysics and cosmology. One avenue to achieve this is direct detection, wherein dark matter in the Milky Way halo interacts with a terrestrial detector, leading to a measurable energy deposit. There is now a highly advanced experimental program dedicated to searching for dark matter at the electroweak scale~\cite{LZ:2022lsv,XENONCollaboration:2022kmb,SuperCDMS:2020aus,SENSEI:2020dpa}. These experiments have made spectacular progress in probing the properties of dark matter, but have not yet revealed a compelling signal. Consequently, there has been much interest in designing new experiments to target unexplored regions of dark matter parameter space, particularly at lower dark matter masses~\cite{Essig:2022dfa}.

Experimental proposals over several decades have considered superfluid helium as a target material~\cite{Lanou:1987eq, Guo:2013dt, Ito:2013cqa, Maris:2017xvi}, with increasing theoretical attention over the past few years~\cite{Schutz:2016tid, Knapen:2016cue, Acanfora:2019con, Caputo:2019cyg, Caputo:2019xum, Baym:2020uos, Caputo:2020sys, Matchev:2021fuw}. Recent experimental proposals have focused on nuclear scattering leading to the quantum evaporation of helium atoms, which can then be detected. There are a number of proposals at various stages of development based on this idea, including HeRALD~\cite{Hertel:2018aal}, DeLIGHT~\cite{vonKrosigk:2022vnf}, ALETHEIA~\cite{Liao:2021npo}, and others~\cite{Lyon:2022sza,You:2022pyn}. These experiments are projected to be sensitive to dark matter masses above $\mathcal{O}(1)$~MeV.

\begin{figure}[t]
    \centering
    \includegraphics[width=0.90\columnwidth]{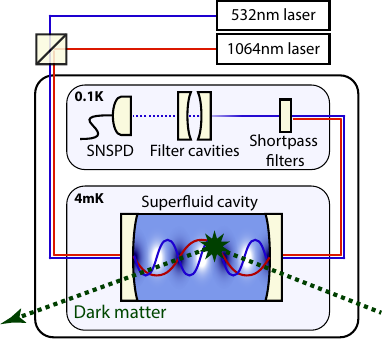}
    \caption{Schematic diagram of the Optomechanical Dark-matter INstrument (ODIN). Dark matter scatters off a highly populated phonon mode (\emph{scattering mode}), which is optically pumped by a 1064~nm laser. The scattered phonon is converted to an anti-Stokes photon through the optomechanical interaction with a $532$\,nm laser. The presence of that photon is registered by a single photon detector after passing through a series of optical filters.}
    \label{fig:detector-schematic}
\end{figure}

At even lower dark matter masses, the relevant target degrees of freedom are no longer helium atoms (or nuclei) but phonons. In this letter, we propose the Optomechanical Dark-matter INstrument (ODIN). ODIN is based on a different detection modality for dark matter scattering on superfluid helium that will be sensitive to this largely unexplored regime using an optomechanical read-out system. The optomechanical interaction enables the transduction of low-energy (undetectable) phonons into high-energy (detectable) photons. With the recent advances of superfluid optomechanics, for example the demonstration of phonon counting with $\mathcal{O}(\mu{\rm eV})$ phonons~\cite{Patil_PRL_22}, we believe our proposal could be implemented with existing technology. This low-energy sensitivity provides access to dark matter in the keV mass range. This is well below the reach of existing experiments, and also probes a very different mass regime from other recent proposals to use optomechanical systems to search for ultralight dark matter~\cite{Carney:2019cio,Manley:2019vxy,Manley_PRL_2021,Brady:2022qne,Murgui:2022zvy}. Another novel feature of ODIN is its ability to modulate the dark matter scattering rate by controlling the phonon density of specific acoustic modes. This could enable lock-in detection, separating the dark matter signal from DC noise sources and providing a pathway to improved sensitivity with fixed target volume.

\section{Optomechanical detection}
\label{sec:optomechanics}

A schematic optomechanical detector is shown in \cref{fig:detector-schematic}. The system is based around a cavity filled with superfluid helium; the cavity simultaneously confines optical and acoustic waves with high spatial overlap~\cite{Kashkanova_NatPhy_2017}. The acoustic waves correspond to density fluctuations in the superfluid (i.e. first sound). Incoming dark matter scatters in the superfluid helium, exciting a phonon. Through the optomechanical interaction with a pumped optical mode, this phonon is converted to an anti-Stokes photon in a resolved sideband~\cite{Kashkanova_NatPhy_2017, Shkarin_PRL_2019, bowen_quantum_2015}. This photon is subsequently detected by a single photon detector after filtering of the optical pump. The system is thus in effect a single-phonon detector. Operating at a temperature of $T\sim4$\,mK, the estimated background rate is $\mathcal{O}(1)$\,event/day.

\begin{figure}[t]
    \centering
    \includegraphics[width=0.9\columnwidth]{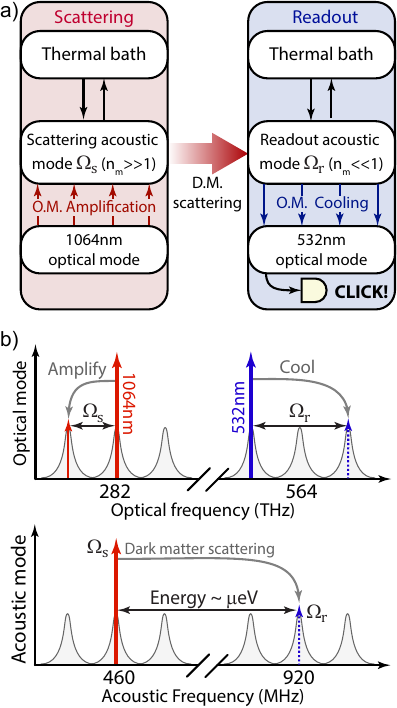}
    \caption{(a) Energy transfer between the two modes. The acoustic scattering mode is populated by optomechanical amplification from the corresponding optical mode. The acoustic readout mode is depopulated by cooling into the corresponding optical mode, read out by the photon detector. In both cases there is also energy transfer with the thermal bath. (b) Energy transfer between the optical modes (top) and mechanical modes (bottom) shown in terms of frequency.}
    \label{fig:optomechanics}
\end{figure}

The optomechanical system described above is a narrow-band detector, sensitive to energy deposition into a specific phonon mode (the \emph{readout mode}) that is determined by the frequency of the optical pump. Scattering dark matter can, however, excite any of the kinematically allowed phonon modes; hence, the scattering rate into the readout mode suffers from a significant phase space suppression. To overcome this we utilise phonon lasing~\cite{Xin_NatPhys_2020} of a second mechanical mode (the \emph{scattering mode}). The dark matter-induced transition between the scattering and readout acoustic modes then benefits from a large Bose enhancement, proportional to the phonon occupation number of the scattering mode. The phonon lasing is achieved by optomechanical amplification via a second pumped optical mode. The detection mechanism is summarised in \cref{fig:optomechanics}a).

The key aspect of the optomechanical interaction is the coherent conversion between phonons and photons, mediated by a strong optical control field. This enables the controllable extraction and deposition of phonons from/to specific mechanical modes. Our detector proposal requires the optomechanical control of two distinct mechanical modes, the \emph{scattering mode} and the \emph{readout mode}, which are amplified and cooled, respectively, through their interaction with pairs of optical cavity modes as shown in \cref{fig:optomechanics}b). The mechanical modes of interest are coupled to specific optical modes via the Brillouin interaction. With this interaction, the photons can induce strain in the superfluid due to electrostriction, and the phonons can scatter light via refractive index modulations due to photoelasticity~\cite{Xin_NatPhys_2020, Kashkanova_NatPhy_2017}. 

Here, we describe the generalised interaction between a single mechanical mode and two optical modes. This formalism applies to both the amplification (red) and cooling (blue) processes shown in \cref{fig:optomechanics}. The optomechanical Hamiltonian that governs the interaction between each mechanical mode and the corresponding optical modes is given by (setting $\hbar=1$)~\cite{Kharel_SciAdv_2019},
\begin{multline}
    H=\omega_2 a^\dagger_2 a_2 + \omega_1 a^\dagger_1 a_1 + \Omega_m b^\dagger_m b_m \\
    - g_0 (a_1 a^\dagger_2 + a_2 a^\dagger_1)(b_m + b_m^\dagger) \,,
\end{multline}
where $a_1$, $a_2$ and $b_m$ are the annihilation operators for the two optical modes and the acoustic mode, with resonance frequencies of $\omega_1$, $\omega_2$, and $\Omega_m$, respectively. To boost the weak single-photon optomechanical coupling rate, $g_0$, we engineer the frequency difference between the optical modes to be equal to the mechanical frequency (i.e. $\Omega_m =\pm(\omega_1-\omega_2)$). This results in a three-mode resonant enhancement of the interaction, reducing the optical power required to control the acoustic mode~\cite{Kharel_SciAdv_2019}. By injecting a strong control field into the $\omega_1$ mode, the relevant interaction Hamiltonian simplifies to 
\begin{equation}
    H_{int} = 
    \begin{cases}
        -g (a^\dagger_2 b_m + b^\dagger_m a_2) \,, & \omega_1 < \omega_2 \,, \\
        -g (a^\dagger_2 b_m^\dagger + b^\dagger_m a_2^\dagger) \,, & \omega_1 > \omega_2 \,,
    \end{cases}
\end{equation}
where $g=\sqrt{N_1}g_0$ is the optomechanical interaction boosted by the number of photons, $N_1$, in the control field.
Under this strong control field, one can derive the Heisenberg-Langevin equations of motion for the dynamics of the photon and phonon modes (see \cref{app:optomechanics} for details). Solving the resulting coupled equations, the rate of phonon amplification/cooling from the optomechanical interaction, in the resolved sideband regime ($\Omega_{m}>\kappa$), is~\cite{Kharel_SciAdv_2019,bowen_quantum_2015}
\begin{equation}
    \Gamma_{\rm om} = \pm\frac{4 g^2}{\kappa} \,,
\end{equation}
where $\kappa$ is the optical loss rate of the cavity. In the case of amplification (i.e. the \emph{scattering mode}), the acoustic mode will experience phonon lasing when $|\Gamma_{\rm om}|>\Gamma_m$, where $\Gamma_m$ is the mechanical loss rate, increasing the phonon occupation until saturated. In the case of cooling (i.e. the \emph{readout mode}), the majority of phonons that enter the system are removed through the optical channel if $\Gamma_{\rm om}>\Gamma_m$.

\section{Dark Matter Scattering Rate}
\label{sec:scattering-rate}

\begin{figure}[ht]
    \centering
    \includegraphics[width=0.95\columnwidth]{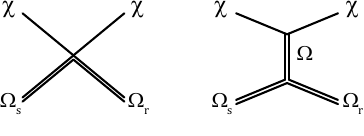}
    \caption{Leading-order contributions to the dark matter--phonon scattering rate. Phonons are represented by double lines.}
    \label{fig:feynman}
\end{figure}
 
We consider spin-independent dark matter--nucleon interactions, which can be expressed as a coupling between dark matter and the number density of the superfluid. The latter can in turn can be written in terms of a phonon field $\phi$, with phonons in the superfluid well-described by an effective field theory~\cite{Son:2002zn}. This yields three- and four- point dark matter--phonon interactions, $\chi\chi\phi$ and $\chi\chi\phi\phi$ respectively, where $\chi$ denotes the dark matter (taken to be a real scalar for illustration). Earlier works~\cite{Schutz:2016tid,Knapen:2016cue,Acanfora:2019con,Caputo:2019cyg,Baym:2020uos} have calculated the rate for dark matter to excite either one or two phonons in bulk superfluid, $\chi\to\chi\phi$ and $\chi\to\chi\phi\phi$. Here, we are interested in the stimulated scattering process $\chi\phi \to \chi\phi$ between phonon modes in the cavity. At leading order in the dark matter--phonon interactions there are two contributions to the scattering rate, shown diagrammatically in \cref{fig:feynman}. Notice that the right diagram proceeds via an intermediate phonon and the anharmonic 3-phonon interaction. When the mediating phonon is on-shell, this contribution is proportional to the acoustic quality factor $Q=\Omega/\Gamma_m$ and provides the dominant contribution to the scattering rate when $Q \gg 1$. 

Neglecting the sub-leading contribution from the left diagram and integrating over the local dark matter velocity distribution, $f_\chi(v)$, the scattering rate is (see \cref{app:EFT,app:phonon-modes,app:scattering-rate} for derivation)
\begin{multline} \label{eq:scattering-rate}
    R = 8\pi^2 \frac{\rho_\chi \sigma_{\chi n}}{m_\chi^3} \big| F_\chi(q) \big|^2 n_s Q^2 \frac{\Omega_r \Omega_s}{m_\mathrm{He}^2 c_s^4} \left(1+\gamma_G\right)^2 \\
    \times \int d^3v\, \left(f_\chi(\vec{v}) + f_\chi(-\vec{v})\right) \delta\Big(\Omega_r-\Omega_s - \Big(q\, \hat{z}\cdot \vec{v} - \frac{q^2}{2m_\chi}\Big)\Big) \,.
\end{multline}
Here, $\rho_\chi$ is the local dark matter density, $m_\chi$ is the dark matter mass, and $\sigma_{\chi n}$ is the dark matter--nucleon scattering cross-section. The number density of phonons in the scattering mode is denoted by $n_s$, the sound speed by $c_s$ and the Gr\"uneisen parameter by $\gamma_G$. The momentum transfer $q=(\Omega_r-\Omega_s)/c_s$ is fixed in terms of the scattering and readout mode energies, $\Omega_s$ and $\Omega_r$ respectively, and is aligned/anti-aligned with the longitudinal direction of the cavity, $\hat{z}$. The dark matter interaction form factor $F_\chi(q)$ is therefore a constant, determined by the reference momentum that defines the cross-section $\sigma_{\chi n}$; henceforth, we fix $F_\chi(q)=1$ without loss of generality.

Notice that, perhaps counter-intuitively, the scattering rate in \cref{eq:scattering-rate} does not scale with the volume of the detector; instead, it is proportional to the number densities of both the scattering phonon mode and the dark matter. This is because the scattering process involves specific initial and final state phonon modes, in contrast to the more familiar situation where one sums over final states within some energy range. 

There is another process, closely related to that calculated above, where dark matter scatters off low-energy, thermally populated phonons instead of the coherent scattering mode. For an experiment running at temperatures of a few mK and with a large phonon occupation in the scattering mode, this contribution to the signal is negligible. 

Finally, while our focus in this letter is superfluid helium, it is worth noting that an analogous two-phonon process will occur in crystals. This raises the possibility of using a highly populated phonon mode to enhance the dark matter interaction rate in proposed solid-state detectors for low-mass dark matter. The relevant scattering rate could be computed using the methods of Ref.~\cite{Campbell-Deem:2019hdx}.

\section{Projected Sensitivity}
\label{sec:projections}

To estimate the achievable sensitivity, we now consider a specific realisation of the ODIN proposal. An important feature of the system is the resonantly enhanced optomechanical interaction, where each acoustic mode is controlled via scattering between two optical modes (see \cref{fig:optomechanics}). The frequencies of the two mechanical modes are $460$\,MHz and $920$\,MHz. These interact with laser control fields that have an optical wavelength equal to twice the acoustic wavelength, corresponding to $1064$\,nm and $532$\,nm, respectively. To ensure the system is triply resonant (i.e. $\omega_1-\omega_2=\Omega_m$) we consider a cavity of length $0.317$\,m, which will exhibit an optical free-spectral-range (FSR) of approximately $460$\,MHz. With this configuration, we expect a single-photon optomechanical coupling rate of $g_{0}/2\pi\approx 1$\,Hz and a cavity loss-rate of $\kappa/2\pi\approx 20$\,kHz when using optical mirrors with 99.99\% reflectivity (see \cref{app:optical_asymmetry,app:optical_loss} for details). Assuming that the intrinsic dissipation of the acoustic mode is limited by 3-phonon processes, as has been observed in Ref.~\cite{Lorenzo_JLTP_2017} and detailed in \cref{app:acoustic_loss}, we predict the Brillouin modes to exhibit $\Gamma_{m}/2\pi\approx10\,\rm Hz$ at $4\,\rm mK$. Injecting $1\rm\,\mu W$ of laser light into the optical resonance leads to a predicted optomechanical amplification/cooling rate of $\Gamma_{\rm om}\approx \pm 24$\,kHz, which is far higher than the rate of thermal processes.

In practice, the cooling rate for the readout mode may need to be reduced to avoid the \emph{strong coupling} regime ($2g>(\kappa,\Gamma)$), which is the optomechanical equivalent of Rabi oscillations and characterised by the coherent exchange between optical and mechanical degrees of freedom~\cite{Verhagen_Nat_2012}. It should also be noted that the density of states of the optical modes must be carefully engineered to suppress the unwanted process of amplification (cooling) of the readout (scattering) mode. One method to achieve this is to insert a thin slab of dielectric into the optical cavity, which would enable six orders of magnitude of suppression (see \cref{app:optical_asymmetry} for details)~\cite{Kharel_SciAdv_2019}. 

Determining the sensitivity also requires a careful estimate of the backgrounds. There are several possible sources, including thermal phonons, dark counts from the single photon detector, and leakage of photons from the control fields of the scattering and readout modes. We discuss each of these in detail in \cref{app:backgrounds}. We argue that dark count rates can be suppressed to $6\times 10^{-6}$\,Hz~\cite{Berggren_PRL22} and that control beam photons can be suppressed with a combination of conventional band-pass filters and cascaded Fabry-Perot cavities. Since essentially every phonon that enters the readout mode is extracted via the optical channel, the thermal phonon occupation must be minimized to limit the background rate. A key feature of this scheme is the use of high frequency acoustic modes which enable exponential suppression of the thermal phonon occupancy when operating at millikelvin temperatures~\cite{Optica_Forstner_2020}. We estimate that an overall background rate of around $10^{-5}$\,Hz is achievable.

\begin{figure}[t]
    \centering
    \includegraphics[width=0.95\columnwidth]{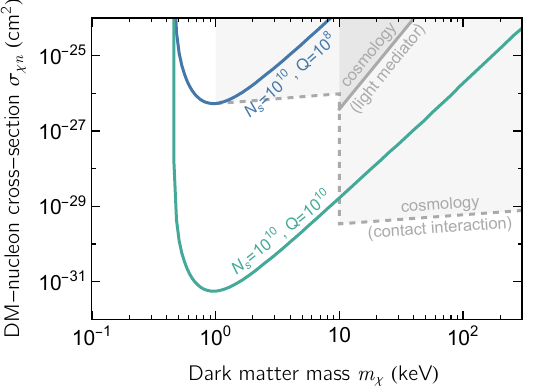}
    \caption{Projected 90\%\,C.L. upper limits on the dark matter--nucleon cross-section at ODIN assuming a run time of 100 days.}
    \label{fig:projected-sensitivity}
\end{figure}

The projected sensitivity to the dark matter--nucleon cross-section $\sigma_{\chi n}$ as a function of dark matter mass is shown in \cref{fig:projected-sensitivity}. Projected 90\% confidence level (C.L.) upper limits were calculated using a Poisson likelihood ratio test statistic, assuming a run time of 100 days. Due to the directionality of the cavity, the signal rate depends on the location and orientation of the detector and will exhibit a daily modulation (assuming the cavity is stationary in the lab frame). In \cref{fig:projected-sensitivity}, we instead show a typical mean rate, obtained by averaging over the orientation of the cavity. We present projections for two different configurations of the experiment, depending on the acoustic quality factor of the cavity, $Q$, that can be achieved. The blue curve in \cref{fig:projected-sensitivity} corresponds to a conservative, \emph{baseline scenario} with $Q=10^8$, noting that Q-factors of $10^8$ have already been demonstrated~\cite{Lorenzo_JLTP_2017}, and a background rate of $10^{-3}$\,Hz. The green curve shows an \emph{improved scenario} with $Q=10^{10}$ and a background rate of $10^{-5}$\,Hz. For both scenarios we assume a fixed phonon population of $N_s=10^{10}$ in the scattering mode (modulation of $N_s$, and hence the dark matter signal, will be explored in future work). These choices are justified in the appendices.
 
The feasibility of operating at $4$\,mK is strongly dependent upon the heat load from optical absorption. For our system parameters, we find that an input optical power of $\sim1$\,nW is required to reach lasing threshold and maintain $N_s=10^{10}$ phonons in the scattering mode (see \cref{app:optical_loss}). Even if $10$\% of this light is unintentionally scattered and absorbed, the resulting heat load of $\sim0.1$\,nW is three orders of magnitude lower than the cooling power of custom dilution refrigerators at $4$\,mK~\cite{Ward_JLTP99}.

There are existing astrophysical and cosmological constraints on $\mathcal{O}(\rm{keV)}$ mass dark matter that interacts with baryons. To compare these with our projected sensitivity we need to specify the structure of the dark matter--baryon interaction. In the case of a contact interaction, the strongest bound is from large scale structure traced by Lyman-$\alpha$~\cite{Rogers:2021byl} for $m_\chi>10$\,keV, with a weaker constraint from the CMB extending down to $m_\chi=1\,$keV~\cite{Gluscevic:2017ywp}. For scattering via a massless mediator, $F_\chi(q)\propto q^{-2}$, the strongest constraint is from CMB+BAO~\cite{Buen-Abad:2021mvc}\footnote{Note that Ref.~\cite{Buen-Abad:2021mvc} constrains the momentum transfer cross-section and there is no model-independent way to translate this to a bound on the direct detection cross-section, as discussed in~\cite{Buen-Abad:2021mvc}. In \cref{fig:projected-sensitivity} we show an approximate bound that assumes the cross-sections are related by a na\"ive $v^{-4}$ scaling.}. These bounds are shown in grey in \cref{fig:projected-sensitivity}. For masses lighter than $m_\chi\sim10$\,keV warm dark matter effects also become relevant and Refs.~\cite{Rogers:2021byl,Buen-Abad:2021mvc} did not calculate bounds in this regime; however, similar constraints are expected to apply. Models of dark matter in this mass range generally feature an additional light particle that mediates the interaction with baryons\footnote{The dark matter would otherwise have been in equilibrium with the baryons at early times and hence in conflict with the upper bound on the number of relativistic degrees of freedom during BBN~\cite{Sabti:2019mhn,Sabti:2021reh}.}. There are strong bounds on such a mediator from stellar cooling, big bang nucleosynthesis, fifth force experiments, and meson decays (see e.g. \cite{Knapen:2017xzo} for a review). These model-dependent bounds do not directly constrain the parameter space in \cref{fig:projected-sensitivity}, but will be relevant when interpreting our projections in the context of specific models. Further discussion of existing bounds on keV--MeV mass dark matter is provided in \cref{app:constraints}.

In the improved scenario (green curve in \cref{fig:projected-sensitivity}), our proposed detector could be sensitive to unexplored parameter space in models with a light mediator. For a dark matter mass of $m_{\chi}=100$\,keV the increase in sensitivity over current cosmological bounds is over three orders of magnitude. For models with a heavy mediator, the improved experiment would explore new parameter space for dark matter masses below $10$\,keV.

Conventional large-scale dark matter detectors typically operate underground to mitigate the cosmogenic background. Here, due to the relatively large cross-sections being considered, an underground experiment would need to carefully consider the shielding impact of the overburden. In the baseline scenario, ODIN would ideally operate in a shallow, sub-surface location; however, future systems with improved sensitivity would need to be deployed in deep underground laboratories. 

\section{Discussion}
\label{sec:discussion}

The use of optomechanical systems is a promising new frontier in the direct detection of dark matter~\cite{Manley:2019vxy,Carney:2020xol,Manley_PRL_2021,Brady:2022qne,Murgui:2022zvy,Monteiro:2020wcb}. Here, we proposed the Optomechanical Dark-matter INstrument (ODIN), a direct detection experiment sensitive to dark matter-nucleon scattering for keV masses. This would provide access to an unexplored mass range, well below the $\gtrsim100$\,MeV probed by current direct detection experiments. ODIN is based on dark matter scattering off a highly populated phonon mode in an optomechanical cavity filled with superfluid helium. Interestingly, the system's sensitivity is primarily dependent on phonon density rather than target volume, in contrast to existing systems. This feature may enable compact, low-cost detectors, with the ability to perform lock-in dark matter detection by periodically depopulating the phonon mode.

ODIN is inherently directional, sensitive only to scattering events where the momentum transfer is aligned with the longitudinal axis of the cavity. A detailed study of the daily modulation of the dark matter signal will therefore be important for a future experiment. Furthermore, it would be interesting to explore alternative detector configurations, possibly involving multiple cavities, to further leverage this directionality.

The methods in this paper may have broader applicability, for example to gravitational wave detection~\cite{Singh:2016xwa,Vadakkumbatt:2021fnw}. It might also be possible to achieve highly populated non-thermal phonon modes in crystals to enhance the scattering rate in solid-state direct detection experiments. We hope to explore these ideas in future work.

\begin{acknowledgments}

The authors would like to thank Ben McAllister for helpful discussions. P.C. and G.H. are supported by Australian Research Council Discovery Early Career Researcher Awards (DE210100446 and DE210100848, respectively). G.H. is also supported by the Big Questions Institute Fellowship. M.J.D. is supported by the Australian Research Council. This work was also supported in part by the Australian Research Council Centre of Excellence for Dark Matter Particle Physics (CE200100008) and Australian Research Council Centre of Excellence for Engineered Quantum Systems (CE170100009).

\end{acknowledgments}


\bibliographystyle{apsrev4-2}
\bibliography{OM_DM}


\appendix

\clearpage
\onecolumngrid

\section{\texorpdfstring{Superfluid $^4\text{He}$ Effective Field Theory}{Superfluid He Effective Field Theory}}
\label{app:EFT}

The relativistic effective field theory (EFT) describing the low-energy phonons in superfluid $^4$He can be written in terms of a real scalar field $\Phi(x)$ as~\cite{Son:2002zn} (see also \cite{Acanfora:2019con,Caputo:2019cyg,Caputo:2019xum,Berezhiani:2020umi,Matchev:2021fuw})
\begin{equation} \label{eq:He4-Lagrangian}
    \mathcal{L} = P(X) \,, \qquad X = \sqrt{\partial^\mu\Phi\partial_\mu\Phi} \,.
\end{equation}
From the corresponding energy-momentum tensor,
\begin{equation}
    T^{\mu\nu} = \frac{P'(X)}{X} \, \partial^\mu\Phi\partial^\nu\Phi - \eta^{\mu\nu} P(X) \,,
\end{equation}
one can identify $P(X)$ as the pressure of the superfluid. The above Lagrangian is invariant under a shift symmetry $\Phi(x) \to \Phi(x) + \alpha$; the conserved charge associated with this symmetry is particle number,
\begin{equation} \label{eq:conserved-charge}
    \int d^3x\, n(x) = \int d^3x\, \frac{P'(X)}{X} \,\partial^0\Phi \,,
\end{equation}
with $n(x)$ the number density. The scalar is assumed to acquire a vacuum expectation value $\langle\Phi(x)\rangle=\mu t$, where $\mu$ is the relativistic chemical potential\footnote{This is related to the standard non-relativistic chemical potential via $\mu_\mathrm{nr} = \mu - m_\mathrm{He}$.}. This breaks both the particle number and time translation symmetries, while preserving the linear combination $H-\mu N$. The average helium number density is $\bar{n} = P'(\mu)$.

Expanding around the background field, we parameterise the phonon (Nambu-Goldstone boson) field, $\phi$, as
\begin{equation}
    \Phi(x,t) = \mu t + \sqrt{\frac{\mu c_s^2}{\bar{n}}}\, \phi(x,t) \,.
\end{equation}
Substituting this into \cref{eq:He4-Lagrangian} yields the low-energy phonon Lagrangian
\begin{equation} \label{eq:Goldstone-Lagrangian}
    \mathcal{L} = \frac{1}{2} \dot\phi^2 - \frac{c_s^2}{2} \left(\nabla\phi\right)^2 + \Big(\frac{1}{2\mu}\left(c_s^2-1\right) + \lambda c_s^2\Big) \sqrt{\frac{\mu c_s^2}{\bar{n}}} \,\dot{\phi}\left(\nabla\phi\right)^2 + \mathcal{O}(\phi^4) \,,
\end{equation}
where the sound speed and cubic coupling are, respectively,
\begin{equation}
    c_s^2 = \frac{P'(\mu)}{\mu P''(\mu)} \,, \qquad
    \lambda = \frac{\mu c_s^2}{6\bar{n}} P'''(\mu) = \frac{1}{6\mu c_s^2}\left(1-2\gamma_G\right) \,,
\end{equation}
with $\gamma_G$ the Gr\"uneisen parameter. Note that in deriving \cref{eq:Goldstone-Lagrangian} we have performed a field redefinition to remove $\dot{\phi}^3$ terms, neglecting operators of $\mathcal{O}(\phi^4)$ or higher in the EFT. The number density can be expressed in terms of the phonon field as
\begin{equation} \label{eq:number-density}
    n(x,t) = \bar{n} + \sqrt{\frac{\bar{n}}{\mu c_s^2}} \dot\phi + \frac{1}{3\mu}\left(\frac{3}{2}c_s^2-1-\gamma_G\right) \left(\nabla\phi\right)^2 + \mathcal{O}(\phi^3) \,.
\end{equation}

\section{Cavity phonon modes}
\label{app:phonon-modes}

The free-field equation of motion for the phonon modes is simply
\begin{equation} \label{eq:eom}
    \ddot\phi - c_s^2\nabla^2\phi = 0 \,,
\end{equation}
with Neumann boundary conditions imposed on the cavity surfaces. Working in cylindrical coordinates $(r,\varphi,z)$, with $z \in[0,L]$, the solution to \cref{eq:eom} for a cylindrical cavity of length $L$ and radius $R$ is
\begin{equation}
    \phi(r,\varphi,z) = \sum_{n=1}^\infty \sum_{k=1}^{\infty} \sum_{m=-\infty}^\infty A_{kmn} \, a_{kmn} \, J_m\left(\frac{j'_{mn} r}{R}\right) \cos\left(\frac{k\pi z}{L}\right) e^{im\varphi} e^{-i\Omega_{kmn} t} + c.c. \,,
\end{equation}
where $k, m, n$ denote the longitudinal, azimuthal, and radial mode numbers, respectively, and $a_{kmn}$ is the annihilation operator for the corresponding mode upon quantisation. $J_m$ is the $m^\text{th}$-order Bessel function, with $j'_{mn}$ its $n^\text{th}$ root. The normalisation constant, $A_{kmn}$, is
\begin{equation}
    A_{kmn} = \sqrt{\frac{2}{\Omega_{kmn}V}} \frac{j'_{mn}}{J_m(j'_{mn}) \sqrt{(j'_{mn})^2-m^2}} \,,
\end{equation}
and the mode energies are given by
\begin{equation}
    \Omega_{kmn} = c_s \sqrt{\left(\frac{j'_{mn}}{R}\right)^2 + \left(\frac{k\pi}{L}\right)^2} \,.
\end{equation}

The quantum numbers for the readout and scattering modes are $(k,m,n)=(k_{r,s},0,1)$, with $k_{r,s}\simeq\Omega_{r,s}L/(\pi c_s)$. We consider a cavity of dimensions $L=31.7$\,cm and $R=0.0725$\,cm, resulting in $k_r=2.4\times10^6$ and $k_s=1.2\times10^6$ for the $\Omega_r=920$\,MHz and $\Omega_s=460$\,MHz modes, respectively. For simplicity, when calculating the scattering rate in the following section, we approximate these cylindrical cavity modes with rectangular cavity modes of the form $\cos(\pi x/R)\cos(\pi y/R)\cos(k_{r,s}\pi z/L)$.

\section{Dark matter -- superfluid scattering rate}
\label{app:scattering-rate}

We consider dark matter that interacts with helium atoms via an interaction of the form
\begin{equation} \label{eq:DM-interaction}
    V_\chi(x) = \sqrt{\pi A^2\sigma_{\chi n}} \, \chi(x)^2 \int d^3x'\, F_\chi(x-x') n(x') \,,
\end{equation}
with $\chi$ the dark matter field (here taken to be a real scalar for simplicity), $\sigma_{\chi n}$ the spin-independent dark matter--nucleon cross-section, $A=4$ the atomic mass number of $^4$He, and $F_\chi(x)$ the position-space dark matter form factor. The helium number density, $n(x)$, is given by \cref{eq:number-density}.

In general, the dark matter event rate in a detector of volume $V$ is
\begin{equation} \label{eq:general-rate}
    R = \frac{\rho_\chi V}{m_\chi} \int d^3v\, f_\chi(\vec{v}) \Gamma_\text{scat}(\vec{v}) \,, 
\end{equation}
with $m_\chi$ the dark matter mass, $\rho_\chi$ the local dark matter density, and $f_\chi(\vec{v})$ the local dark matter velocity distribution. For the latter, we assume a truncated Maxwell-Boltzmann distribution and use the astrophysical parameters recommended in \cite{Baxter:2021pqo}. In the Born approximation for an interaction of the form \eqref{eq:DM-interaction}, the scattering rate as a function of the dark matter velocity is given by
\begin{equation}
    \Gamma_\text{scat}(\vec{v}) = \frac{\pi A^2\sigma_{\chi n}}{m_\chi^2 V} \int \frac{d^3q}{(2\pi)^3}\, 2\pi \delta\left(\Omega_r - \Omega_s -\Omega(\vec{q},\vec{v}) \right) \big| F_\chi(q) \big|^2 \, \big| \matrixel{\Omega_r}{n(q)}{n_s,\Omega_s} \big|^2 \,, 
\end{equation}
where $\Omega(\vec{q},\vec{v})=\vec{q} \cdot \vec{v} - q^2/(2m_\chi)$ is the energy transfer, $n_s$ is the number density of the scattering phonon mode, and $\Omega_s$, $\Omega_r$ are the energies of the scattering and readout phonons. The Fourier transforms of the dark matter form factor and helium number density are denoted by $F_\chi(q)$ and $n(q)$, respectively.

We are interested in the regime $N_s \equiv n_s V \gg 1$, such that the scattering rate benefits from a large Bose enhancement. Working to first order in the phonon self-interactions there are two contributions to the scattering process, as shown in Fig. 3 (main text). The right diagram is resonantly enhanced as the intermediate phonon can be on-shell. Consequently, it provides the dominant contribution to the scattering rate in the narrow width limit, i.e. when $Q=\Omega/\Gamma_\Omega \gg 1$, with $\Gamma_\Omega$ the phonon decay width. In this limit, we obtain
\begin{equation} \label{eq:Gamma-rate}
    \Gamma_\text{scat}(\vec{v}) = \frac{\pi^2}{2V} \frac{A^2\sigma_{\chi n}}{m_\chi^2} \big| F_\chi(q) \big|^2 n_s \frac{\Omega_r \Omega_s}{\mu^2c_s^4} \left(1+\gamma_G-\frac{3}{2}c_s^2\right)^2 \left(\frac{\Omega}{\Gamma_\Omega} \right)^2 \delta\Big(\Omega_r-\Omega_s - \Big(q\, \hat{z}\cdot \vec{v} - \frac{q^2}{2m_\chi}\Big)\Big) + (\hat{z} \to -\hat{z}) \,,
\end{equation}
where the momentum transfer is fixed by the scattering and readout modes to have magnitude $q=(\Omega_r-\Omega_s)/c_s$ and be either aligned or anti-aligned with the longitudinal direction of the cavity, $\hat{z}$. In deriving the above expression we have assumed a linear phonon dispersion, which is a very good approximation for the relevant phonon energies. Substituting \cref{eq:Gamma-rate} into \cref{eq:general-rate} yields the total scattering rate in Eq.~(4) of the main text. The measured values for the sound speed and Gr\"uneisen parameter at $T<0.1$\,K and saturated vapour pressure are $c_s=238.2$\,m/s~\cite{Donnelly:1998} and $\gamma_G=2.84$~\cite{Abraham:1970}.

\section{\texorpdfstring{Optomechanical interaction with superfluid $^4\text{He}$ Brillouin mode}{Optomechanical interaction with superfluid He Brillouin mode}}
\label{app:optomechanics}

In keeping with the convention used by the optomechanics community, here we do not set $\hbar=1$ (unlike in the main text). The optomechanical Hamiltonian describing a single mechanical mode coupled to a pair of optical modes, with one optical mode driven by an external field, is given by Ref.~\cite{Kharel_SciAdv_2019},
\begin{equation}
    H=\hbar\omega_1 a^\dagger_1 a_1 +\hbar\omega_2 a^\dagger_2 a_2 +\hbar\Omega_m b^\dagger_m b_m \\
    - \hbar g_0(a^\dagger_2 a_1 + a^\dagger_1 a_2)(b_m + b^\dagger_m) + i\hbar \sqrt{\kappa_{ext}} \alpha_{1,in} (a^\dagger_{1,in} e^{-i\omega_{1,in} t}-a_{1,in} e^{i\omega_{1,in} t}) \,,
\end{equation}
where $\omega_1$ and $\omega_2$ are the resonance frequencies of two optical modes, $\omega_{1,in}$ is the frequency of the drive, $\Omega_m$ is the mechanical resonance, $g_0$ is the optomechanical single-photon coupling rate, and $\kappa_{ext}$ is the coupling rate in/out of the cavity as determined by the reflectivity of each mirror. The total loss rate of the cavity is given by $\kappa=2\kappa_{ext}+\kappa_0$, where $\kappa_0$ is the intrinsic absorption within the optical cavity (here $\kappa_0\ll\kappa_{ext}$). The power input into the cavity is determined by $P_{in}=\hbar \omega_1 |\alpha_{1,in}|^2$. Typically the single-photon coupling rate is low, requiring a bright control beam to boost the interaction~\cite{bowen_quantum_2015}. This can be further enhanced by ensuring the cavity modes are separated by the mechanical resonance frequency ($\omega_2-\omega_1=\Omega_m$). First, we consider the situation of optomechanical cooling. Assuming we are driving with a bright coherent beam on one optical mode (i.e. $\omega_{1,in}=\omega_{1}$) and monitoring the higher energy optical mode, we can then linearise the Hamiltonian to obtain
\begin{equation}
    H=\hbar\omega_2 a^\dagger_2 a_2 +\hbar\Omega_m b^\dagger_m b_m - \hbar g_0 \sqrt{N_1}(a^\dagger_2 b_m + b^\dagger_m a_2) \,,
\end{equation}
where $N_1$ is the number of intra-cavity photons given the launched optical power. Operating on resonance, this is $N_1=\frac{2}{\kappa}\frac{P_{in}}{\hbar \omega_1}$.  
From this Hamiltonian one can derive the following Heisenberg-Langevin equations of motion, in the rotating frame, for the dynamics of the anti-Stokes photon and the phonon modes,
\begin{align}
    \Dot{b}_{m}&=\left (i\Omega_m-\frac{\Gamma_m}{2}\right )b_{m}(t)+ig_0 \sqrt{N_1}a_{2} +\sqrt{\Gamma_{m}}b_{in} \,, \\
    \Dot{a}_{2}&=\left (i\Omega_m-\frac{\kappa}{2}\right )a_{2}(t)+ig_0 \sqrt{N_1}b_{m} +\sqrt{\kappa_{ext}}a_{in} \,.
\end{align}
These equations can then be solved to find the modification to the mechanical dissipation rate. This corresponds to the rate of phonon extraction via the optomechanical interaction and is given by,
\begin{equation}
    \Gamma_{\rm om} = \frac{4 N_1 g_0^2}{\kappa} \,.
    \label{eq:Gamma-rate-SI}
\end{equation}

For the situation of optomechanical cooling, if $\Gamma_{\rm om}$ is much higher than the intrinsic dissipation rate, then the majority of phonons that enter the system are up-converted to photons and subsequently removed through the optical channel. Alternatively, to perform optomechanical amplification, we can drive the higher frequency optical mode with the control field, which manifests as a sign change in Eq.~\ref{eq:Gamma-rate-SI}. Both of these processes depend on the strength of the optomechanical coupling which, for Brillouin modes, is given by~\cite{Kashkanova_NatPhy_2017, Shkarin_thesis, Harris_Optica_2020}, 
\begin{equation*}
    g_{0}=\omega\frac{(n_{He}^2-1)}{2n_{He}^2} \sqrt{\frac{\hbar\Omega_m}{2 V K}}\eta \,,
\end{equation*}
where $\omega$ is the optical frequency, $n_{He}$ is the refractive index, $K$ is the bulk modulus of superfluid helium, $V$ is the effective mode volume, and $\eta$ is the effective mode overlap. For the geometry of the cavity considered here, we estimate an optomechanical coupling rate of $g_{0}/2\pi\approx 1\rm Hz$.
Using the expected $g_0$ and a cavity loss rate of $\kappa/2\pi \approx 20\,\rm{kHz}$, with an input of $1\rm\, \mu W$ of laser light, then $\Gamma_{\rm om}/2\pi\approx 24\, \rm kHz$ (as compared to the intrinsic dissipation $\Gamma_{\rm m}/2\pi\approx10\,\rm Hz$). 

\begin{figure}[ht]
    \centering
    \includegraphics[width=0.75\textwidth]{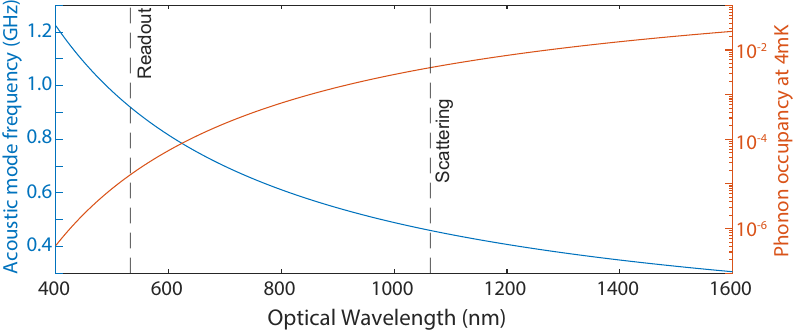}
    \caption{Relationship between the optical wavelength $\lambda$, the corresponding Brillouin mode frequency $\Omega=2\pi c_s/\lambda$ (left axis), and the phonon occupation of the Brillouin mode at $4\,\rm mK$ (right axis).}
\end{figure}

\section{Acoustic Loss} 
\label{app:acoustic_loss}

At low temperatures ($T < 600 \,\rm mK$) the main intrinsic loss mechanism for density waves in superfluid helium is via a three phonon process~\cite{MarisMassey,Kashkanova_NatPhy_2017}. The intrinsic quality factor $Q$ for a mode of frequency $\Omega$ and width $\Gamma$ can be shown, in the $T\to 0$ limit, to be
\begin{equation}
    Q =\frac{\Omega}{\Gamma} = \frac{240\pi}{(1+\gamma_G)^2} \frac{\rho_{\rm He} c_s^5}{\hbar\Omega^4} \,,
\end{equation}
where $\rho_{\rm He}$ is the density of the superfluid. For the relevant acoustic mode frequency $\Omega=2\pi\times460\, \rm MHz$ this gives $Q=8\times10^{11}$.

At these low temperatures the acoustic quality factor may be dominated by extrinsic processes, for example, acoustic radiation into the confining mirrors. This can be ameliorated by implementing acoustic Distributed Bragg Reflectors (DBRs) to minimize acoustic radiation. This method of attaining high quality factors in superfluid helium is detailed in the supplementary section of Ref.~\cite{Kashkanova_NatPhy_2017}. Acoustic quality factors of the order of $Q\approx10^8$ have been observed in bulk density waves, albeit at lower frequencies, a result which was limited by dilute $^3$He impurities~\cite{Lorenzo_JLTP_2017}.

\section{Phonon occupancy}
\label{app:phonon_occupancy}

Optomechanical systems can exhibit ``phonon-lasing'' when the gain from the optomechanical interaction surpasses the intrinsic acoustic dissipation. When entering this regime the phonon occupancy will rise exponentially until nonlinearities saturate the gain, which fundamentally limits the ultimate phonon occupancy. Assuming that the leading nonlinearity arises from the optomechanical interaction itself, the dispersive shift $G$ of the cavity per unit strain $\epsilon$ (where $\epsilon=\delta V/V=\delta \rho/\rho$  and $\rho$ is density) is given by:
\textbf{\begin{equation}
    G =\frac{g_0}{\epsilon_{zpf}} = \frac{g_0}{\sqrt{\hbar\Omega/(K V)}} \,,
\end{equation}}
where $g_{0}$ is the single-photon optomechanical coupling rate, $V$ is the effective mode volume of the acoustic mode, $K$ is the bulk modulus of the superfluid helium, and $\epsilon_{zpf}$ is the zero-point volumetric strain of the acoustic mode~\cite{Harris_Optica_2020}.

As the phonon occupancy grows, the dispersive shift of the optical cavity increases, eventually resulting in a shift that is larger than the optical linewidth. This results in a detuning of the pump laser and a reduction of the optomechanical driving, occurring at the saturation amplitude of  
\textbf{\begin{equation}
    \epsilon_{sat} =\frac{\kappa}{G} \,,
\end{equation}}
where $\kappa$ is the optical linewidth. We can then estimate the corresponding phonon occupation $N_{ph}$ by calculating the energy contained within the acoustic mode when oscillating at the saturation amplitude,
\textbf{\begin{equation}
     \hbar \Omega N_{ph} = \frac{1}{2}K V \epsilon^2_{sat} \,.
\end{equation}}
The resulting upper bound on the phonon occupancy depends only on the ratio of the optical linewidth to the single-photon optomechanical coupling, 
\textbf{\begin{equation}
     N_{ph} = \left(\frac{\kappa}{2 g_0}\right)^2 \,,
\end{equation}}
leading to the desire for large optical dissipation and low optomechanical coupling.

Taking an optomechanical coupling rate of $g_{0}/2\pi\approx 1\rm Hz$ and cavity linewidth of $\kappa/2\pi\approx 20\, \rm kHz$ (corresponding to 99.99\% reflectivity mirrors), we find the phonon occupancy is limited to $N_{ph}\approx10^8$. If lower reflectivity mirrors are used and $\kappa/2\pi\approx 200\,\rm kHz$ (corresponding to 99.9\% reflectivity mirrors), the phonon occupancy is limited to $N_{ph}\approx10^{10}$. For reference, phonon occupancies of $N_{ph}\approx 10^{12}$ have been observed in solid state Brillouin systems~\cite{Kharel_SciAdv_2019}.

\section{Optical loss}
\label{app:optical_loss}

The optical absorption through helium is immeasurably low so that all of the optical loss/absorption comes from the mirrors. The quality of these mirrors affects the optomechanical coupling rate (through $\kappa$) and the minimum attainable temperature (through heating). The quality factor of a Fabry-Perot resonator can be calculated from the reflectivity of the mirrors ($T_{1}$ and $T_{2}$), the cavity length ($L$), and the frequency of light ($f$):
\begin{equation}
    Q = \frac{f}{\kappa} = \frac{4\pi f L}{c (T_{1}+T_{2})} \,.
\end{equation}

Taking as an example crystalline mirrors offered by Thorlabs (XM12R8), which have reflectivity of $99.99\%$, and incorporating them into a $31.7\,\rm cm$ long cavity, we obtain a linewidth of $\kappa \approx 20\,\rm kHz$. Importantly, the absorption is specified to be $<1\,\mathrm{ppm}$ which means a pump beam as high as $1\,\rm mW$ would correspond to an exceptionally low heating of $1\,\rm nW$. For comparison, commercial dilution fridges can achieve $14-30\,\rm \mu W$ of cooling power at $20\,\rm mK$~\cite{DilFridgeReview} and custom dilution fridges can achieve $300\,$nW of cooling power at $4\,$mK~\cite{Ward_JLTP99}. It should be noted that the mirrors considered in this proposal will need to have a dual-band coating to ensure high reflectivity at both $1064\,$nm and $532\,$nm. Dual band mirrors at these wavelengths are commercially available with reflectivities above $99.5\%$ (for example, Edmund Optics Dual Band Laser Line \#20-371).

We now address in detail the sensitivity of our experiment to the relevant heat loads. The heat load from optical absorption of the $1064$\,nm beam can be estimated by calculating the photon flux required to reach lasing threshold and then maintain a large phonon occupation in the scattering mode. A phonon occupation of $N_{ph}\approx10^{10}$ results in phonons leaving the system at a rate of $N_{ph}\Gamma_m =3\times 10^{9}\,\text{s}^{-1}$ (with $Q\approx10^{10}$ corresponding to  $\Gamma_m/2\pi \approx 0.046$\,Hz for the $460$\,MHz mode). If these phonons are to be replaced via the optomechanical interaction then we must have a comparable number of photons entering the cavity since each photon generates of the order of one phonon (See section S6A of supplemental information in Ref.~\cite{Kharel_SciAdv_2019}). A photon flux of $3\times 10^{9}\,\text{s}^{-1}$ corresponds to an input optical power of $\sim1$\,nW at $1064$\,nm. The absorption in crystalline mirrors can be as low as $1$\,ppm (Thorlabs XM12R8), however, some light could be scattered from impurities or misalignment and absorbed elsewhere in the cryostat. Even if $10$\% of the light is scattered and absorbed, the resulting heat load is $\sim0.1$\,nW, which is three orders of magnitude lower than the cooling power of custom dilution refrigerators at $4$\,mK~\cite{Ward_JLTP99}, and thus experimentally accessible. Furthermore, it should be noted that the filter cavities, which absorb the pump beams before the SNSPD, are thermally anchored to a separate cooling stage within the cryostat (the ``$100$\,mK'' stage), which typically has $>500\,\mu$W of cooling power.

\section{Optical Asymmetry}
\label{app:optical_asymmetry}

\begin{figure}[ht]
    \centering
    \includegraphics[width=0.90\textwidth]{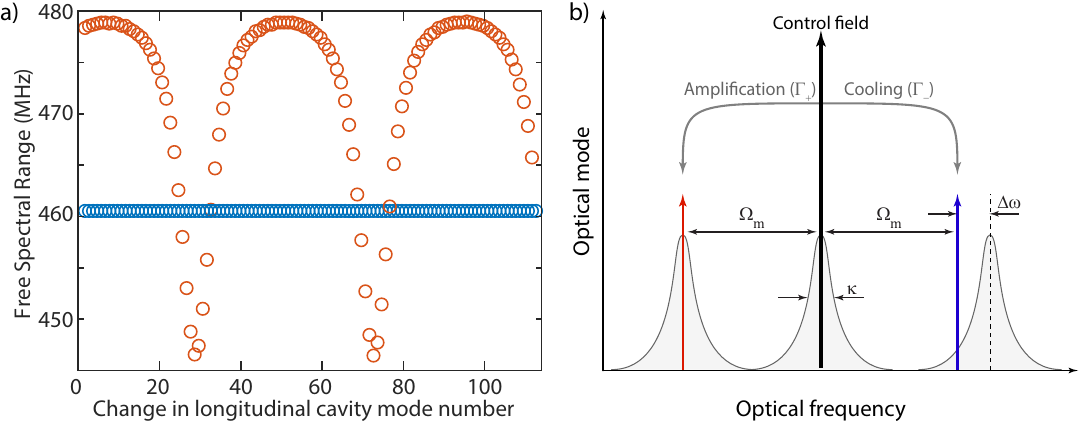}
    \caption{(a) Spectral separation between longitudinal modes of a $31.7\,$cm long optical cavity, i.e. the free spectral range (FSR). The bare cavity (blue circles) shows a constant FSR, whereas the cavity with a 5\,mm slab of silicon dioxide (orange circles) shows variation in the FSR. This spectral asymmetry can be used to suppress unwanted optomechanical interactions. (b) Optical cavity mode spectrum with a 5\,mm slab of silicon dioxide. Due to the asymmetric FSR, the cooling process is resonantly suppressed while the amplification process is enhanced.}
 \label{fig:FSR}
\end{figure}

To achieve efficient cooling (amplification) for the readout (scattering) beam, we must suppress the amplification (cooling) scattering process. In a single-mode optomechanical system, this is typically done by detuning the pump beam from the optical resonance (for $\Omega_m \gg \kappa$). For the multi-mode system considered here, the coherent driving field is directly on resonance with an optical mode. The asymmetry in amplification/cooling can only arise from a different spectral separation between subsequent optical modes. This spectral separation is often referred to as the free-spectral-range (FSR). We can engineer the FSR by adding a slab of dielectric material into the cavity, which acts as an etalon that modulates the effective length of the cavity. The FSR of a regular $31.7\,$cm cavity (blue dots), and one with a $5\,$mm slab of silicon dioxide placed at one end (orange circles), is shown in Fig~\ref{fig:FSR}. Considering a $460\,$MHz Brillouin acoustic mode, we see that for the cavity with the dielectric slab only some optical mode pairs are resonant. This results in a cavity enhancement of one scattering process and a cavity suppression of the other. 

We can explicitly calculate the difference in amplification/cooling rate by extending the input-output formalism used in Section \ref{app:optomechanics}. This calculation is detailed in the supplementary information of Ref.~\cite{Kharel_SciAdv_2019}, with the result
\begin{equation}
    \Gamma_{-}=\Gamma_{+}\left(\frac{\kappa}{2 \Delta \omega}\right)^2\approx \Gamma_{+} 10^{-6} \,,
\end{equation}
where $\Gamma_{-}$ is the cooling rate, $\Gamma_{+}$ is the amplification rate, $\kappa$ is the optical linewidth and $\Delta \omega$ is the additional detuning for the non-resonant process. For our system parameters (i.e. $10\rm \,kHz$ optical linewidth and $5\,\rm MHz$ detuning), we find the unwanted amplification/cooling process can be suppressed by $10^{6}$.

\section{Single Photon Background Rate}
\label{app:backgrounds}

The experimental platform proposed here relies upon the detection of single photons to herald a dark matter event. It is therefore crucial to achieve  a detailed understanding of the possible backgrounds. Here we review the various sources of these spurious signals and bound their values. We argue that it should be possible to suppress the background rate to $10^{-5}$~Hz.

\subsection{Dark Count Rate}

The dark count rate (DCR) of the single photon detector will be a limiting factor in the achievable sensitivity of our experiment. Superconducting nanowire single photon detectors (SNSPD) typically offer the lowest DCR with the highest detection efficiency. It has been shown that isolated SNSPDs (i.e. not fiber-coupled out of the cryostat) can exhibit a DCR of $6\times10^{-6}$\,Hz~\cite{Berggren_PRL22}. However, when  fibre-coupling these SNSPDs out of the cryostat to room temperature, the DCR increases by many orders-of-magnitude to values as large as $1$\,Hz. This massive increase arises from thermal photons produced at room temperature being guided down cladding modes of the fiber optic cable and directed onto the SNSPD. This source of DCR can be effectively mitigated by filtering incoming light with  cryogenic narrow-band filters~\cite{Tokura_OptLett2015}. With this level of cryogenic filtering a DCR of $~6\times10^{-6}$\,Hz will be achievable.

\subsection{Filtering pump beam}

We must heavily filter the light leaving the superfluid cavity to ensure no stray photons from both the cooling beam (from the readout mode) and the amplification beam (from the scattering mode) arrive at the SNSPD. As we propose to use $1\,\mu$W of pump light, we need greater than $200$\,dB of suppression to achieve a leakage rate of $1$ photon per day. The filtering stages will need to be cryogenic free-space optics; this keeps the optical loss to a minimum and prevents the generation of thermal photons.

Filtering of the heating beam at $1064$\,nm can be achieved with a stack of conventional band-pass filters, which typically have a bandwidth of $1$\,nm, an optical depth (OD) of 6-7, and transmission of greater than $95$\,\%. For example, three Edmund Optics Ultra Narrow Filters (\#36-640, $532$\,nm, OD6-7) would likely provide sufficient suppression. Filtering of the cooling beam is more challenging since the signal photons (i.e. those from dark matter interaction) are only $1$\,GHz shifted from the bright carrier. This filtering must be done with cascaded Fabry-Perot filter cavities. For example, three cascaded cavities with FSR of $3$\,GHz and finesse of $10^4$ will likely provide sufficient suppression.

\subsection{Thermal occupation of acoustic mode}

The thermal occupancy of the readout acoustic mode needs to be kept as low as possible. Dilution cryostats are widely used in quantum information and communication applications and can, with customization, operate continuously at $4$\,mK with $>300$\,nW of cooling power \cite{Ward_JLTP99}. At this temperature, the thermal occupation for an acoustic mode of $920\,\rm MHz$ is
\begin{equation}
    \langle b^{\dagger}_{m}b_{m}\rangle=\frac{1}{\exp{\hbar \Omega /k_{b}T}-1}\approx 10^{-5} \,.
    \label{EqBoseEinstein}
\end{equation}

The rate of phonon heating from the bath is therefore $\langle b^{\dagger}_{m}b_{m}\rangle\Gamma\approx 10^{-3}\,\rm phonon/s$ for a mode with an acoustic  quality factor of $10^8$. In the case of large optomechanical cooling ($\Gamma_{\rm om} \gg\Gamma_m$, see section~\ref{app:optomechanics}), as considered here, essentially every phonon that enters the system, either from the bath or from dark matter interactions, is pulled out through the optical channel. The mode temperature thus provides a source of background noise for dark matter detection.

\section{Existing constraints on keV-MeV mass dark matter}
\label{app:constraints}

In this section we briefly review the existing constraints on keV--MeV mass dark matter that interacts with baryons. First, recall that our proposal is sensitive to dark matter scattering at a specific momentum transfer, $q=(\Omega_r-\Omega_s)/c_s\sim$\,eV. To compare our projected sensitivity with existing bounds therefore requires an additional assumption about the energy-dependence of the dark matter--baryon interaction, which is parameterised by the dark matter form factor $F_\chi(q)$. Two limiting cases of interest are $F_\chi(q)=\text{const}$ and $F_\chi(q)\propto q^{-2}$; the former corresponds to a point-like, contact-interaction and the latter to an interaction mediated by a massless particle (or a particle with mass much smaller than the momentum transfer, $q$, in all processes of interest). Fig.~4 of the main text shows the leading, model-independent\footnote{Some model-dependence enters in the $F_\chi(q)\propto q^{-2}$ case when translating from the momentum transfer cross-section to the direct detection cross-section, see.~\cite{Buen-Abad:2021mvc}.} bounds from cosmology for these two cases, which are due to modifications of large scale structure as probed by Lyman-$\alpha$~\cite{Rogers:2021byl} (contact interaction) and the cosmic microwave background~\cite{Buen-Abad:2021mvc} (light mediator).

It is well known that the dark matter baryon--interaction cannot, however, be described by a contact interaction up to arbitrarily high energies. If the interaction were to remain point-like at energies above that of neutrino decoupling ($\gtrsim2$\,MeV), then cross-sections $\sigma_{\chi n} \gtrsim 10^{-31}\,\text{cm}^2$ are large enough that the dark matter would be in equilibrium with the photons at this time. The dark matter then contributes significantly to the energy density of the universe during big bang nucleosynthesis (BBN), modifying the abundances of the light elements. Consequently, dark matter with mass $\lesssim$~MeV that was in equilibrium at early times is excluded~\cite{Sabti:2019mhn,Sabti:2021reh}. However, if the dark matter-baryon interaction is mediated by a particle with mass $\lesssim$~MeV, the dark matter can be out of equilibrium in the early universe and still yield a low-energy scattering cross-section that is potentially observable.

Our proposal is therefore predominantly sensitive to models which feature a light (sub-MeV) mediator that couples to baryons. There are strong constraints on such mediators from a variety of sources, including stellar cooling~\cite{Hardy:2016kme,Yamamoto:2023zlu}, SN1987A~\cite{Dev:2020eam}, BBN~\cite{Sabti:2019mhn,Sabti:2021reh}, fifth force experiments~\cite{Banks:2020gpu} and meson decays (e.g. $K \to \pi+\text{invisible}$~\cite{NA62:2021zjw}). For mediators in the eV\,--\,100\,keV mass range, the strongest bounds are due to energy loss in red giant or horizontal branch stars, while for sub-eV mediator masses fifth force experiments provide the leading constraint. Dark matter self-interactions also place an upper bound on the coupling of the dark matter to the mediator (see e.g. Ref.~\cite{Tulin:2017ara}). Ref.~\cite{Knapen:2017xzo} studied the combined impact of these constraints in a minimal model with a scalar mediator that couples to either gluons or top quarks. For that particular model, requiring that these existing bounds are satisfied constrains the direct detection cross-section to be below the projected reach of ODIN, assuming the model accounts for 100\% of the dark matter abundance. If the model provides only a sub-component of the dark matter, relaxing the bound from dark matter self-interactions, that might allow for an observable signal. This particular example highlights the importance of considering constraints on the mediator when assessing the potential sensitivity of direct detection experiments to low mass dark matter; however, the constraints and resulting conclusions are necessarily model-dependent and require a dedicated analysis for each model of interest.

Finally, while the scattering of non-relativistic keV--MeV mass dark matter is well below the reach of existing direct detection experiments, they are potentially sensitive to a flux of (semi-)relativistic dark matter. Such a flux can be produced in the up-scattering of dark matter by cosmic rays~\cite{Bringmann:2018cvk}, or from other astrophysical sources. This could also produce an observable signal in large neutrino detectors such as Borexino or Super-Kamiokande. However, the resulting bounds can again only be interpreted within the context of a specific model and, as recently shown in Ref.~\cite{Bell:2023sdq}, bounds on the mediator (as discussed above) generally exclude the region that can be probed via cosmic ray up-scattering.

\end{document}